\documentclass[preprint]{aastex631}

\begin{document}

\title{The gravitational redshift of solar-type stars from {\it Gaia} DR3 wide binaries}

\author{Kareem El-Badry}
\affiliation{Center for Astrophysics $|$ Harvard \& Smithsonian, 60 Garden Street, Cambridge, MA 02138, USA}
\affiliation{Harvard Society of Fellows, 78 Mount Auburn Street, Cambridge, MA 02138}
\affiliation{Max-Planck Institute for Astronomy, K\"onigstuhl 17, D-69117 Heidelberg, Germany}

\begin{abstract}
Light escaping from a gravitational potential suffers a redshift with magnitude proportional to the depth of the potential. This ``gravitational redshift'' is easily measurable in dense stars such as white dwarfs, but is much weaker and has evaded unambiguous detection in main-sequence stars.  I show that the effect is directly measurable in the {\it Gaia} DR3 radial velocities (RVs) of the components of wide binary stars. In a sample of $\sim$500 wide binaries containing a solar-type main-sequence star and a red giant or red clump companion, the apparent RV of the giant is on average $0.49 \pm 0.02 \,\, \rm km\,s^{-1}$ lower than that of the main-sequence star. 
This owes primarily to the giants' weaker gravitational fields and is in reasonably good agreement with the value expected from general relativity. 
\end{abstract}



\section{Introduction} \label{sec:intro}
General relativity predicts that light emitted from the surface of a massive object undergoes a redshift, mimicking the Doppler shift when light is emitted from a receding object. For a star of mass $M$ and radius $R$, the amplitude of the induced apparent RV in the weak-field limit is 
\begin{equation}
\label{eq:vgr}
v_{\rm GR}=\frac{GM}{Rc}=0.64\,{\rm km\,s^{-1}}\left(\frac{M}{M_{\odot}}\right)\left(\frac{R}{R_{\odot}}\right)^{-1}.
\end{equation}
This velocity is large and easily detectable in white dwarfs, where values of about $30\,\rm km\,s^{-1}$ are typical \citep[e.g.][]{Falcon2010}. In neutron stars, it is enormous \citep[of order $0.3c$; e.g.][]{Cottam2002}. In main-sequence stars, it is smaller, but still comfortably within the capabilities of modern spectrographs.

An obvious challenge in measuring gravitational redshifts is that stars are not stationary, but have quasi-random velocities along our line of sight, with typical amplitudes of tens of $\rm km\,s^{-1}$. This stymies detection of weak gravitational redshifts unless the star's true radial velocity can be determined through external information. Wide binary stars offer precisely the required external information: their orbital velocities are small, such that to first approximation, the two stars in a wide binary have the same RV, and any difference between their {\it observed} RVs can be attributed to gravitational redshift. This allowed \citet{Greenstein1967} and many subsequent works to measure gravitational redshifts and masses of white dwarfs in wide binaries with main-sequence companions. 

The situation is more complicated for binaries containing two main-sequence stars, because (a) the gravitational redshifts of both components are expected to have similar magnitude, and (b) the orbital velocity and surface convective shift both have comparable magnitude to the expected gravitational redshift. This can complicate attempts to precisely measure genuine RV differences in wide binaries \citep[e.g.][]{Loeb2022}, but it also suggests that the effects of gravitational redshift should be detectable in a large sample of wide binaries. This has recently become possible thanks to the {\it Gaia} mission, which both enables selection of pure wide binaries samples using astrometry, and measures RVs for the component stars.

The most straightforward approach is to compare the RVs of giants and dwarfs, as the giants are expected to have much smaller (usually negligible) redshifts.  \citet[][]{Pasquini2011} compared the mean RVs of a large sample of dwarfs and giants in a star cluster, where the peculiar velocities should average out. However, they were unable to detect the gravitational redshift statistically, attributing the non-detection to the confounding effects of surface convection. More recently, \citet[][]{Gutierrez2022} did detect a gravitational redshift signal in the RVs of star cluster members, but only when combining data from many different clusters and fitting a global model. Similarly, \citet[][]{Moschella2022} recently found tentative evidence of gravitational redshift in main-sequence wide binaries selected from {\it Gaia} DR2 data, but the effect was not obvious because {\it Gaia} DR2 contained RVs for both components of only a few giant/dwarf binaries.

{\it Gaia} DR3 \citep[][]{Valenari2022} includes precise RVs to stars $\approx$ 2 magnitudes fainter than DR2, leading to a much larger usable sample of giant/dwarf binaries. Here, I leverage these new RVs to measure a visually obvious signature of gravitational redshift in the velocity difference of wide binaries' components. 

\section{Methods}

From the wide binary catalog constructed by \citet[][]{El-Badry2021}, I selected binaries satisfying the following cuts:

\begin{itemize}
    \item The position of the primary (the brighter component) in the color-magnitude diagram suggests it is a giant or red clump star, satisfying $M_{G} < 2\times(G_{\rm BP}-G_{\rm RP}) -1$. I corrected for extinction using the 3D dust map of \citet[][]{Green2019} in the north and that of \citet[][]{Lallement2019} in the south.
    \item The position of the secondary suggest it is a dwarf, with $M_{G} > 2\times(G_{\rm BP}-G_{\rm RP}) + 0.5$.
    \item The pair is likely to be gravitationally bound, with a chance alignment probability below 10\% (\texttt{R\_chance\_align} $<$ 0.1). In practice, almost all the selected binaries have \texttt{R\_chance\_align} $<$ 0.01.
    \item Both stars have a RV measurement reported in {\it Gaia} DR3, with \texttt{radial\_velocity\_error} $<2\,\rm km\,s^{-1}$.
    \item The pair is separated by at least 5 arcsec, to minimize the effects of blending on the RVs \citep{Katz2022}. The median projected physical separation of the sample is 6,000\,AU, corresponding to a typical orbital RV difference of order $0.5\,\rm km\,s^{-1}$ (with no preferred sign).
\end{itemize}

I next ``corrected'' the {\it Gaia} RVs using the empirical magnitude-dependent RV zeropoint derived by \citet[][their equation 5]{Katz2022}. The magnitude of this correction is $< 0.1\,\rm km\,s^{-1}$ for a majority of the sample. 

\begin{figure*}
    \centering
    \includegraphics[width=\textwidth]{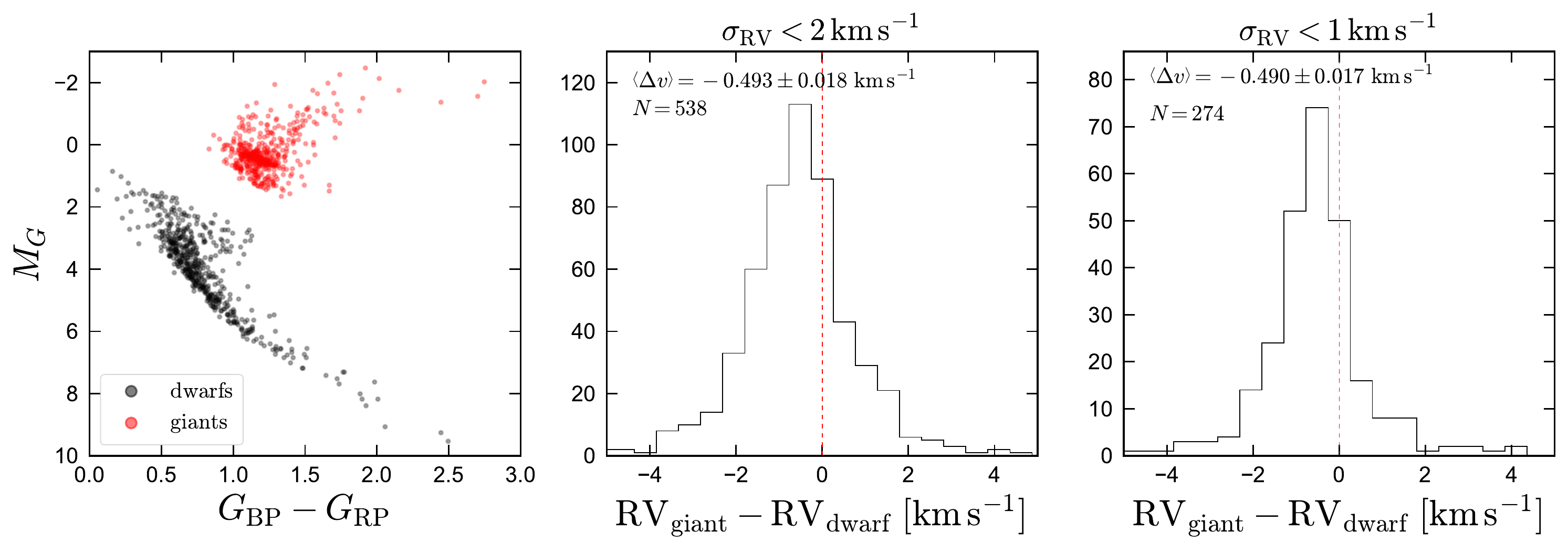}
    \caption{Left: color-magnitude diagram of the cleaned sample of 538 giant/dwarf wide binaries. Most of the dwarfs are near the main-sequence with mass $0.5\lesssim M/M_{\odot} \lesssim 1.5$; most of the ``giants'' are core helium burning red clump stars. Center: radial velocity difference between the giant and dwarf for this full sample. Right: radial velocity difference for the subset in which both components have radial velocity uncertianties below $1\,\rm km\,s^{-1}$. The fact that the distributions are not centered on 0 owes to the larger gravitational redshift in the dwarfs. }
    \label{fig:grav}
\end{figure*}

These cuts yielded 538 binaries, which are shown on the color-magnitude diagram in the left panel of Figure~\ref{fig:grav}. The median apparent magnitude of the giants is $G\approx 8.3$, and that of the dwarfs is $G\approx 12.0$. The dwarfs are mostly solar-type stars with masses $0.5\lesssim M/M_{\odot} \lesssim 1.5$. A majority are near the main sequence, but a few subgiants are also included in the sample. A majority of the ``giants'' are not strictly on the giant branch but are core helium burning red clump stars, with $R\approx 10\,R_{\odot}$ and (most likely)  $ M \approx 1\,R_{\odot}$. The median distance to the sample is 350 pc. 

The main result is shown the center and right panels of Figure~\ref{fig:grav}: there is a not-very-subtle difference in the mean RV of dwarfs and giants. The inverse-variance weighted mean RV difference is $\approx 0.49\,\rm km\,s^{-1}$, both for the full sample (center column) and for the subsample in which both components have \texttt{radial\_velocity\_error} $<1\,\rm km\,s^{-1}$ (right column). The observed RV difference distribution is narrower for latter subsample because the RV differences for individual binaries are still dominated by measurement uncertainties. 

The expected RV difference due to gravitational redshift is ${\rm \Delta}v_{{\rm GR}}=\frac{G}{c}\left(M_{{\rm giant}}/R_{{\rm giant}}-M_{{\rm dwarf}}/R_{{\rm dwarf}}\right)$. Assuming $R_{{\rm dwarf}} \approx R_{\odot}\left(M_{{\rm dwarf}}/M_{\odot}\right)$, $M_{\rm giant}\approx 1\,M_{\odot}$, and $R_{\rm giant}\approx 10\,R_{\odot}$, this translates to an expected typical ${\rm \Delta}v_{{\rm GR}}\approx 0.58\,\rm km\,s^{-1}$, pleasingly close to the observed value. 

\section{Conclusion}
I have shown that the different gravitational redshifts of main-sequence and giant stars leads to an unambiguous shift in their apparent relative RVs in wide binaries observed by {\it Gaia}. I have not attempted to account for star-to-star variation in $M/R$ within the dwarf and giant samples, or for the effects of convective shifts that can lead the mean RV of the photosphere to differ from that of a star's center of mass. The fact that the observed mean RV offset is close to the value predicted due to gravitational redshift suggests that these simplifications are tolerable at the $\lesssim 20\%$ level, at least in a population-averaged sense, but a more detailed investigation is certainly warranted. 

Given the uncertainties in RV calibration and the still poorly understood effects of convective shifts, this approach seems unlikely to yield precise tests of GR in the near future. It is, however, a nice demonstration of the gravitational redshift phenomenon and of the precision and stability of the {\it Gaia} DR3 RV data.

\section*{Acknowledgements}
I thank George Seabroke and David  W. Hogg for discussions that prompted this note. This work was done at the {\it Gaia Hike} meeting at the University of British Columbia in June 2022. This work has made use of data from the European Space Agency (ESA) mission Gaia (https://www.cosmos.esa.int/gaia), processed by the Gaia Data Processing and Analysis Consortium (DPAC, https://www.cosmos.esa.int/web/gaia/dpac/consortium). Funding for the DPAC has been provided by national institutions, in particular the institutions participating in the Gaia Multilateral Agreement.

\bibliographystyle{aasjournal}

\end{document}